\documentclass[pdflatex,sn-aps]{sn-jnl}

\usepackage{setspace}
\newcommand*{\ExtComp}{1}   

\usepackage[utf8]{inputenc} 
\ifdefined\HelvFont
\usepackage{helvet}
\fi
\usepackage{geometry}			
\usepackage{hyperref} 		
\usepackage[nomain,acronym,toc]{glossaries}
\usepackage{graphicx}
\usepackage{tabularx}
\usepackage{color}				
\usepackage[dvipsnames]{xcolor}				
\usepackage{amsfonts}
\usepackage{amsthm}
\usepackage{fontenc} 
\usepackage{textcomp} 
\usepackage{braket} 
\usepackage{wasysym}
\usepackage{mathtools}
\usepackage{dcolumn} 
\usepackage{lmodern} 			
\usepackage{adjustbox}
\usepackage{multirow}
\usepackage{pgfplots}			
\usepackage{tikz}					
\usetikzlibrary{shapes}
\usetikzlibrary{calc}
\usetikzlibrary{positioning}
\usepgfplotslibrary{fillbetween}
\usepgfplotslibrary{colormaps} 
\usetikzlibrary{shapes.geometric}
\usetikzlibrary{arrows}
\usepackage{circuitikz}
\usepackage{tikzscale}		
\usepackage{latexsym} 		
\usepackage{float}				
\usepackage{fancyhdr} 		
\usepackage{titlesec} 		
\usepackage{listings}
\usepackage{marginnote}
\usepackage[normalem]{ulem}
\usepackage{etoolbox}
\usepackage{appendix}
\usepackage{makecell}
\usepackage{relsize}

\usepackage[eulergreek]{sansmath}

\ifdefined\ExtComp
\usepgfplotslibrary{external}
\fi

\ifdefined\HelvFont

\renewcommand{\rmdefault}{\sfdefault}
\usepackage{mathastext}
\pgfplotsset{every tick label/.append style={font={\sffamily}}}
\pgfplotsset{every node/.style={/utils/exec={\sfdefault}}}
\fi

\definecolor{naranja_1}{rgb}{0.70,0.25,0}
\definecolor{gris-azul}{rgb}{0.27,0.33,0.415}  
\definecolor{azul-gris}{rgb}{0.192,0.278,0.4}  
\definecolor{verde-chulo}{rgb}{0.25,0.5,0.5}  
\definecolor{verde-chulo2}{rgb}{0,0.5,0.4} 

\definecolor{color1}{rgb}{1, 0, 0}%
\definecolor{color2}{rgb}{0, 0, 1}%
\definecolor{color3}{rgb}{0, 1, 0}%
\definecolor{color4}{rgb}{1, 0, 1}%
\definecolor{color5}{rgb}{0, 1, 1}%

\definecolor{mycolor11}{RGB}{166,189,219}%
\definecolor{mycolor6}{RGB}{116,169,207}%
\definecolor{mycolor1}{RGB}{54,144,192}%
\definecolor{mycolor16}{RGB}{5,112,176}%
\definecolor{mycolor21}{RGB}{4,90,141}%
\definecolor{mycolor17}{RGB}{252, 146, 114}%
\definecolor{mycolor7}{RGB}{251, 106, 74}%
\definecolor{mycolor12}{RGB}{239, 59, 44}%
\definecolor{mycolor2}{RGB}{203, 24, 29}%
\definecolor{mycolor22}{RGB}{165, 15, 21}%
\definecolor{mycolor18}{RGB}{250, 159, 181}%
\definecolor{mycolor8}{RGB}{247, 104, 161}%
\definecolor{mycolor13}{RGB}{221, 52, 151}%
\definecolor{mycolor5}{RGB}{174, 1, 126}%
\definecolor{mycolor23}{RGB}{122, 1, 119}%
\definecolor{mycolor19}{RGB}{254,196,79}%
\definecolor{mycolor9}{RGB}{254,153,41}%
\definecolor{mycolor14}{RGB}{236,112,20}%
\definecolor{mycolor4}{RGB}{204,76,2}%
\definecolor{mycolor24}{RGB}{153,52,4}%
\definecolor{mycolor20}{RGB}{153, 216, 201}%
\definecolor{mycolor10}{RGB}{102, 194, 164}%
\definecolor{mycolor15}{RGB}{65, 174, 118}%
\definecolor{mycolor3}{RGB}{35, 139, 69}%
\definecolor{mycolor25}{RGB}{0, 109, 44}%

\pgfplotscreateplotcyclelist{molecColorList}{
color1, line width=0.8pt\\
color2, line width=0.8pt\\
color3, line width=0.8pt\\
color4, line width=0.8pt\\
color5, line width=0.8pt\\
color1, mark=o\\
color2, mark=o\\
color3, mark=o\\
color4, mark=o\\
color5, mark=o\\
}

\pgfplotsset{%
	colormap={parula}{
	rgb255=(53,42,135)
	rgb255=(15,92,221)
	rgb255=(18,125,216)
	rgb255=(7,156,207)
	rgb255=(21,177,180)
	rgb255=(89,189,140)
	rgb255=(165,190,107)
	rgb255=(225,185,82)
	rgb255=(252,206,46)
	rgb255=(249,251,14)
	},
	colormap/bluered,
	colormap/jet,
	colormap={RdBu}{
	rgb255=(103,1,31)
	rgb255=(176,23,42)
	rgb255=(213,95,26)
	rgb255=(243,163,128)
	rgb255=(253,218,198)
	rgb255=(255,255,255)
	rgb255=(209,229,240)
	rgb255=(144,196,221)
	rgb255=(66,146,194)		
	rgb255=(32,100,170)		
	rgb255=(5,48,167)
	},
	colormap={linBR}{
	rgb255=(0,0,255)
	rgb255=(27,0,228)
	rgb255=(54,0,201)
	rgb255=(80,0,175)
	rgb255=(107,0,148)
	rgb255=(134,0,121)
	rgb255=(161,0,94)
	rgb255=(188,0,67)
	rgb255=(215,0,40)
	rgb255=(241,0,14)
	rgb255=(255,0,0)
	},
	colormap={linBR2}{
	rgb255=(0,0,255)
	rgb255=(54,0,201)
	rgb255=(107,0,148)
	rgb255=(161,0,94)
	rgb255=(215,0,40)
	rgb255=(255,0,0)
	},
	colormap={linBR3}{
	rgb255=(0,0,255)
	rgb255=(126,0,129)
	rgb255=(255,0,0)
	},
	colormap={linBRmain}{
		rgb255=(0,0,255)
		rgb255=(14,0,241)%
		rgb255=(27,0,228)
		rgb255=(40,0,215)%
		rgb255=(54,0,201)
		rgb255=(67,0,188)%
		rgb255=(80,0,175)
		rgb255=(94,0,161)%
		rgb255=(107,0,148)
		rgb255=(121,0,134)%
		rgb255=(134,0,121)%
		rgb255=(148,0,107)
		rgb255=(161,0,94)%
		rgb255=(175,0,80)
		rgb255=(188,0,67)%
		rgb255=(201,0,54)
		rgb255=(215,0,40)%
		rgb255=(228,0,27)
		rgb255=(241,0,14)%
		rgb255=(255,0,0)
	},
	colormap={cmap1}{
		rgb255=(53,42,135)
		rgb255=(55,47,148)
		rgb255=(55,53,160)
		rgb255=(55,58,171)
		rgb255=(53,62,180)
		rgb255=(51,67,189)
		rgb255=(47,71,196)
		rgb255=(43,76,203)
		rgb255=(38,80,208)
		rgb255=(32,84,213)
		rgb255=(25,88,217)
		rgb255=(16,92,221)
		rgb255=(5,95,223)
		rgb255=(0,99,224)
		rgb255=(0,102,225)
		rgb255=(0,105,224)
		rgb255=(0,108,223)
		rgb255=(0,111,222)
		rgb255=(4,113,221)
		rgb255=(9,116,219)
		rgb255=(14,119,218)
		rgb255=(17,122,217)
		rgb255=(18,124,216)
		rgb255=(18,127,216)
		rgb255=(17,130,215)
		rgb255=(16,133,215)
		rgb255=(15,136,214)
		rgb255=(14,139,214)
		rgb255=(12,142,213)
		rgb255=(11,145,212)
		rgb255=(9,148,212)
		rgb255=(8,150,210)
		rgb255=(7,153,209)
		rgb255=(7,155,208)
		rgb255=(7,158,206)
		rgb255=(5,160,204)
		rgb255=(2,162,202)
		rgb255=(0,164,200)
		rgb255=(0,166,197)
		rgb255=(0,168,195)
		rgb255=(0,170,192)
		rgb255=(0,172,190)
		rgb255=(0,173,187)
		rgb255=(7,175,184)
		rgb255=(17,176,181)
		rgb255=(26,178,178)
		rgb255=(33,179,175)
		rgb255=(39,181,172)
		rgb255=(44,182,168)
		rgb255=(50,183,164)
		rgb255=(56,184,160)
		rgb255=(56,184,160)
		rgb255=(85,189,142)
		rgb255=(121,190,126)
		rgb255=(155,190,112)
		rgb255=(185,188,99)
		rgb255=(213,185,88)
		rgb255=(233,187,76)
		rgb255=(245,196,60)
		rgb255=(253,209,42)
		rgb255=(255,228,24)
		rgb255=(249,251,14)
	},
	colormap={linBRmain100}{
		rgb255=(0,0,255)
		rgb255=(2,0,253)
		rgb255=(4,0,251)
		rgb255=(6,0,249)
		rgb255=(8,0,247)
		rgb255=(10,0,245)
		rgb255=(12,0,243)
		rgb255=(14,0,241)
		rgb255=(16,0,239)
		rgb255=(18,0,237)
		rgb255=(20,0,235)
		rgb255=(22,0,233)
		rgb255=(24,0,231)
		rgb255=(26,0,229)
		rgb255=(28,0,227)
		rgb255=(30,0,225)
		rgb255=(32,0,223)
		rgb255=(34,0,221)
		rgb255=(36,0,219)
		rgb255=(38,0,217)
		rgb255=(40,0,215)
		rgb255=(42,0,213)
		rgb255=(44,0,211)
		rgb255=(46,0,209)
		rgb255=(48,0,207)
		rgb255=(50,0,205)
		rgb255=(52,0,203)
		rgb255=(54,0,201)
		rgb255=(56,0,199)
		rgb255=(58,0,197)
		rgb255=(60,0,195)
		rgb255=(62,0,193)
		rgb255=(64,0,191)
		rgb255=(66,0,189)
		rgb255=(68,0,187)
		rgb255=(70,0,185)
		rgb255=(72,0,183)
		rgb255=(74,0,181)
		rgb255=(76,0,179)
		rgb255=(78,0,177)
		rgb255=(80,0,175)
		rgb255=(82,0,173)
		rgb255=(84,0,171)
		rgb255=(86,0,169)
		rgb255=(88,0,167)
		rgb255=(90,0,165)
		rgb255=(92,0,163)
		rgb255=(94,0,161)
		rgb255=(96,0,159)
		rgb255=(98,0,157)
		rgb255=(100,0,155)
		rgb255=(102,0,153)
		rgb255=(104,0,151)
		rgb255=(106,0,149)
		rgb255=(108,0,147)
		rgb255=(110,0,145)
		rgb255=(112,0,143)
		rgb255=(114,0,141)
		rgb255=(116,0,139)
		rgb255=(118,0,137)
		rgb255=(120,0,135)
		rgb255=(122,0,133)
		rgb255=(124,0,131)
		rgb255=(126,0,129)
		rgb255=(128,0,127)
		rgb255=(130,0,125)
		rgb255=(132,0,123)
		rgb255=(134,0,121)
		rgb255=(136,0,119)
		rgb255=(138,0,117)
		rgb255=(140,0,115)
		rgb255=(142,0,113)
		rgb255=(144,0,111)
		rgb255=(146,0,109)
		rgb255=(148,0,107)
		rgb255=(150,0,105)
		rgb255=(152,0,103)
		rgb255=(154,0,101)
		rgb255=(156,0,99)
		rgb255=(158,0,97)
		rgb255=(160,0,95)
		rgb255=(162,0,93)
		rgb255=(164,0,91)
		rgb255=(166,0,89)
		rgb255=(168,0,87)
		rgb255=(170,0,85)
		rgb255=(172,0,83)
		rgb255=(174,0,81)
		rgb255=(176,0,79)
		rgb255=(178,0,77)
		rgb255=(180,0,75)
		rgb255=(182,0,73)
		rgb255=(184,0,71)
		rgb255=(186,0,69)
		rgb255=(188,0,67)
		rgb255=(190,0,65)
		rgb255=(192,0,63)
		rgb255=(194,0,61)
		rgb255=(196,0,59)
		rgb255=(198,0,57)
		rgb255=(200,0,55)
		rgb255=(202,0,53)
		rgb255=(204,0,51)
		rgb255=(206,0,49)
		rgb255=(208,0,47)
		rgb255=(210,0,45)
		rgb255=(212,0,43)
		rgb255=(214,0,41)
		rgb255=(216,0,39)
		rgb255=(218,0,37)
		rgb255=(220,0,35)
		rgb255=(222,0,33)
		rgb255=(224,0,31)
		rgb255=(226,0,29)
		rgb255=(228,0,27)
		rgb255=(230,0,25)
		rgb255=(232,0,23)
		rgb255=(234,0,21)
		rgb255=(236,0,19)
		rgb255=(238,0,17)
		rgb255=(240,0,15)
		rgb255=(242,0,13)
		rgb255=(244,0,11)
		rgb255=(246,0,9)
		rgb255=(248,0,7)
		rgb255=(250,0,5)
		rgb255=(252,0,3)
		rgb255=(254,0,1)
	},
	colormap={linBR9}{
		rgb255=(0,0,255)
		rgb255=(28,0,227)
		rgb255=(56,0,199)
		rgb255=(84,0,171)
		rgb255=(112,0,143)
		rgb255=(140,0,115)
		rgb255=(168,0,87)
		rgb255=(196,0,59)
		rgb255=(224,0,31)
		rgb255=(252,0,3)
	},
	colormap={linBR5}{
		rgb255=(0,0,255)
		rgb255=(51,0,204)
		rgb255=(102,0,153)
		rgb255=(153,0,102)
		rgb255=(204,0,51)
	},
	colormap={linBR21}{	
	rgb255=(0,0,255)
	rgb255=(12,0,243)
	rgb255=(24,0,231)
	rgb255=(36,0,219)
	rgb255=(48,0,207)
	rgb255=(60,0,195)
	rgb255=(72,0,183)
	rgb255=(84,0,171)
	rgb255=(96,0,159)
	rgb255=(108,0,147)
	rgb255=(120,0,135)
	rgb255=(132,0,123)
	rgb255=(144,0,111)
	rgb255=(156,0,99)
	rgb255=(168,0,87)
	rgb255=(180,0,75)
	rgb255=(192,0,63)
	rgb255=(204,0,51)
	rgb255=(216,0,39)
	rgb255=(228,0,27)
	rgb255=(240,0,15)
	rgb255=(252,0,3)
	},
	colormap={newBWR12}{
	rgb255=(0,0,255)
	rgb255=(51,51,255)
	rgb255=(102,102,255)
	rgb255=(153,153,255)
	rgb255=(204,204,255)
	rgb255=(255,255,255)
	rgb255=(255,204,204)
	rgb255=(255,153,153)
	rgb255=(255,102,102)
	rgb255=(255,51,51)
	rgb255=(255,0,0)
	},
	colormap={BlueScale3}{
	rgb255=(34, 108, 166)
	rgb255=(72, 145, 200)
	rgb255=(132,186, 224)
	rgb255=(0,39,82)
	},
	colormap={natMap30}{
		rgb255=(0,39,82)
		rgb255=(13,65,113)
		rgb255=(26,91,145)
		rgb255=(43,125,186)
		rgb255=(64,139,176)
		rgb255=(92,157,164)
		rgb255=(113,170,154)
		rgb255=(137,186,147)
		rgb255=(156,197,142)
		rgb255=(177,210,142)
		rgb255=(189,215,154)
		rgb255=(201,221,165)
		rgb255=(215,227,178)
		rgb255=(226,231,187)
		rgb255=(239,237,197)
		rgb255=(245,232,192)
		rgb255=(245,215,168)
		rgb255=(245,202,149)
		rgb255=(245,186,125)
		rgb255=(245,173,107)
		rgb255=(245,161,88)
		rgb255=(238,141,76)
		rgb255=(231,126,71)
		rgb255=(222,105,64)
		rgb255=(219,91,68)
		rgb255=(215,74,74)
		rgb255=(212,61,79)
		rgb255=(174,48,58)
		rgb255=(146,38,41)
		rgb255=(118,28,25)
	},
	colormap={natBlueRed30}{
		rgb255=(0,39,82)
		rgb255=(13,51,91)
		rgb255=(26,63,101)
		rgb255=(44,79,114)
		rgb255=(57,91,123)
		rgb255=(75,107,136)
		rgb255=(88,119,146)
		rgb255=(106,135,159)
		rgb255=(120,147,168)
		rgb255=(137,163,181)
		rgb255=(151,175,191)
		rgb255=(164,187,200)
		rgb255=(182,203,213)
		rgb255=(195,215,223)
		rgb255=(213,231,236)
		rgb255=(219,231,234)
		rgb255=(218,216,218)
		rgb255=(217,204,206)
		rgb255=(216,188,190)
		rgb255=(215,176,178)
		rgb255=(215,163,166)
		rgb255=(214,147,150)
		rgb255=(213,135,138)
		rgb255=(212,120,122)
		rgb255=(211,107,110)
		rgb255=(210,92,94)
		rgb255=(209,79,82)
		rgb255=(208,64,66)
		rgb255=(207,51,54)
		rgb255=(207,40,42)
	},
	colormap={natBlueRedA30}{
		rgb255=(0,39,82)
		rgb255=(18,55,95)
		rgb255=(36,71,108)
		rgb255=(60,93,125)
		rgb255=(78,110,138)
		rgb255=(102,131,156)
		rgb255=(120,148,169)
		rgb255=(145,170,186)
		rgb255=(163,186,199)
		rgb255=(187,208,217)
		rgb255=(205,224,230)
		rgb255=(219,236,239)
		rgb255=(219,223,226)
		rgb255=(218,213,216)
		rgb255=(217,201,204)
		rgb255=(216,191,194)
		rgb255=(216,179,181)
		rgb255=(215,169,172)
		rgb255=(214,157,159)
		rgb255=(214,147,150)
		rgb255=(213,138,140)
		rgb255=(212,125,127)
		rgb255=(211,115,118)
		rgb255=(211,103,105)
		rgb255=(210,93,96)
		rgb255=(209,81,83)
		rgb255=(209,71,73)
		rgb255=(208,58,61)
		rgb255=(207,49,51)
		rgb255=(207,40,42)
	},
	colormap={natBlueRedB30}{
		rgb255=(207,40,42)
		rgb255=(209,82,84)
		rgb255=(212,125,127)
		rgb255=(216,182,184)
		rgb255=(219,224,227)
		rgb255=(212,230,235)
		rgb255=(204,223,229)
		rgb255=(193,214,222)
		rgb255=(186,207,216)
		rgb255=(175,198,209)
		rgb255=(168,191,203)
		rgb255=(160,184,197)
		rgb255=(150,174,190)
		rgb255=(142,167,184)
		rgb255=(131,158,177)
		rgb255=(124,151,171)
		rgb255=(113,141,164)
		rgb255=(106,134,158)
		rgb255=(95,125,151)
		rgb255=(87,118,145)
		rgb255=(80,111,139)
		rgb255=(69,102,132)
		rgb255=(62,95,126)
		rgb255=(51,85,119)
		rgb255=(43,78,113)
		rgb255=(33,69,106)
		rgb255=(25,62,100)
		rgb255=(15,53,93)
		rgb255=(7,46,87)
		rgb255=(0,39,82)
	},
	colormap={natBlueRedC30}{
		rgb255=(0,39,82)
		rgb255=(9,47,88)
		rgb255=(19,56,95)
		rgb255=(32,68,105)
		rgb255=(41,76,112)
		rgb255=(54,88,121)
		rgb255=(64,97,128)
		rgb255=(76,108,137)
		rgb255=(86,117,144)
		rgb255=(99,128,153)
		rgb255=(109,137,160)
		rgb255=(118,146,167)
		rgb255=(131,157,177)
		rgb255=(141,166,183)
		rgb255=(153,178,193)
		rgb255=(163,186,200)
		rgb255=(176,198,209)
		rgb255=(186,207,216)
		rgb255=(198,218,225)
		rgb255=(208,227,232)
		rgb255=(218,236,239)
		rgb255=(218,215,218)
		rgb255=(217,196,199)
		rgb255=(215,170,172)
		rgb255=(214,150,153)
		rgb255=(212,124,127)
		rgb255=(211,105,107)
		rgb255=(209,79,81)
		rgb255=(208,59,61)
		rgb255=(207,40,42)
	},
	colormap={natBlueRedD30}{
		rgb255=(0,39,82)
		rgb255=(25,62,100)
		rgb255=(51,85,119)
		rgb255=(86,117,144)
		rgb255=(112,140,163)
		rgb255=(147,172,188)
		rgb255=(173,195,207)
		rgb255=(207,226,232)
		rgb255=(219,233,236)
		rgb255=(219,222,225)
		rgb255=(218,214,217)
		rgb255=(217,206,209)
		rgb255=(217,196,198)
		rgb255=(216,187,190)
		rgb255=(216,177,179)
		rgb255=(215,169,171)
		rgb255=(214,158,160)
		rgb255=(214,150,152)
		rgb255=(213,139,142)
		rgb255=(213,131,133)
		rgb255=(212,123,125)
		rgb255=(211,112,115)
		rgb255=(211,104,106)
		rgb255=(210,93,96)
		rgb255=(210,85,87)
		rgb255=(209,74,77)
		rgb255=(208,66,69)
		rgb255=(208,56,58)
		rgb255=(207,48,50)
		rgb255=(207,40,42)
	},
	colormap={natBlueRedE30}{
		rgb255=(0,39,82)
		rgb255=(17,54,94)
		rgb255=(34,70,107)
		rgb255=(57,91,123)
		rgb255=(75,106,136)
		rgb255=(98,127,152)
		rgb255=(115,143,165)
		rgb255=(138,164,182)
		rgb255=(156,180,194)
		rgb255=(179,201,211)
		rgb255=(196,216,223)
		rgb255=(213,232,236)
		rgb255=(219,228,231)
		rgb255=(218,218,221)
		rgb255=(217,205,208)
		rgb255=(217,195,198)
		rgb255=(216,182,185)
		rgb255=(215,173,175)
		rgb255=(214,160,162)
		rgb255=(214,150,153)
		rgb255=(213,140,143)
		rgb255=(212,127,130)
		rgb255=(212,117,120)
		rgb255=(211,104,107)
		rgb255=(210,95,97)
		rgb255=(209,82,84)
		rgb255=(209,72,74)
		rgb255=(208,59,61)
		rgb255=(207,49,51)
		rgb255=(207,40,42)
	},
	}

\pgfplotsset{compat=1.14}
\pgfplotsset{minor grid style={dashed}}
\pgfplotsset{
	cycle list/.define={linBR}{[of colormap=linBRmain]},
	colormap={linBR}{
		indices of colormap=
		(0,2,...,\pgfplotscolormaplastindexof{linBRmain} of linBRmain)
	}, 
}
\pgfplotsset{
	cycle list/.define={linBR}{[of colormap=linBRmain]},
	colormap={linBRrev}{
		indices of colormap=(
			\pgfplotscolormaplastindexof{linBRmain},...,0 of linBRmain
		)
	},
	cycle list/.define={linBRrev}{[of colormap=linBRrev]},
}
\pgfplotsset{
	colormap={linBR7rev}{
		indices of colormap=(
			0,3,...,\pgfplotscolormaplastindexof{linBRrev} of linBRrev
		)
	},
	cycle list/.define={linBR7rev}{[of colormap=linBR7rev]},
}
\pgfplotsset{
	colormap={linBR7}{
		indices of colormap=(
			0,3,...,\pgfplotscolormaplastindexof{linBRmain} of linBRmain
		)
	},
	cycle list/.define={linBR7}{[of colormap=linBR7]},
}
\pgfplotsset{
	colormap={linBR6rev}{
		indices of colormap=(
			0,3,...,\pgfplotscolormaplastindexof{linBRrev} of linBRrev
		)
	},
	cycle list/.define={linBR6rev}{[of colormap=linBR6rev]},
}
\pgfplotsset{
	colormap={linBR6}{
		indices of colormap=(
			0,3,...,\pgfplotscolormaplastindexof{linBRmain} of linBRmain
		)
	},
	cycle list/.define={linBR6}{[of colormap=linBR6]},%
}

\hypersetup{
	colorlinks = true,
	linkcolor = azul-gris,
	citecolor = naranja_1,
	urlcolor = orange,
	runcolor = black,
	breaklinks = true,
}

\newcommand{\myName}{A. Toral-Lopez}

\titleformat{name=\chapter, numberless}											
[display]																									
{\bfseries\large}																					
{
}
{-7ex}																										
{\centering\MakeUppercase}
[\vspace{5ex}]					
\titlespacing*{name=\chapter, numberless}{0pt}{120pt}{6pt}

\ifdefined\SingleRep
	\titleformat{\chapter}[display]
	{\centering \bfseries \Large} 
	{}
	{0ex}
	{}
	[\vspace{0.1cm} \centering \large \sf \myName \vspace{0.2cm} \rule{1.0\textwidth}{0.5pt}] 
	\titlespacing*{\chapter}{0pt}{-20pt}{12pt}
\else
	\titleformat{\chapter}[display]
	{\centering \bfseries \huge}
	{}
	{0ex}
	{}
	[\rule{1.0\textwidth}{0.5pt}]
	\titlespacing*{\chapter}{0pt}{-20pt}{12pt}
\fi

\renewcommand{\thesection}{\arabic{section}}

\fancypagestyle{plain}{
	\fancyhead{}
	
}

\newcommand{\secref}[1]{\hyperref[#1]{Section \ref*{#1}}}
\newcommand{\appref}[1]{\hyperref[#1]{Appendix \ref*{#1}}}
\newcommand{\chref}[1]{\hyperref[#1]{Chapter \ref*{#1}}}
\newcommand{\figref}[1]{\hyperref[#1]{Figure \ref*{#1}}}
\newcommand{\sfigref}[2]{\hyperref[#1]{Figure \ref*{#1}#2}}
\newcommand{\nfigref}[2]{\hyperref[#1]{\ref*{#1}#2}}
\newcommand{\tabref}[1]{\hyperref[#1]{Table \ref*{#1}}}

\DeclareUnicodeCharacter{2212}{-}

\tikzstyle{textFil} = [text opacity=1, fill opacity=0.6]
\tikzstyle{startstop} = [rectangle, rounded corners, minimum width=3cm, minimum height=1cm,text centered, draw=black, fill=red!30]
\tikzstyle{io} = [trapezium, trapezium left angle=70, trapezium right angle=110, minimum width=3cm, minimum height=1cm, text centered, draw=black, fill=blue!30]
\tikzstyle{process} = [rectangle, minimum width=3cm, minimum height=1cm, text centered, draw=black, fill=orange!30]
\tikzstyle{decision} = [diamond, minimum width=3cm, minimum height=1cm, text centered, draw=black, fill=green!30]
\tikzstyle{arrow} = [thick, ->, >=stealth]

\pgfkeys{/tikz/savenumber/.code 2 args={\global\edef#1{#2}}}

\newcolumntype{L}[1]{>{\raggedright\arraybackslash}p{#1}}
\newcolumntype{C}[1]{>{\centering\arraybackslash}p{#1}}
\newcolumntype{R}[1]{>{\raggedleft\arraybackslash}p{#1}}

\makeatletter
\def\clearpage{%
	\ifvmode
	\ifnum \@dbltopnum =\m@ne
	\ifdim \pagetotal <\topskip
	\hbox{}
	\fi
	\fi
	\fi
	\newpage
	\thispagestyle{empty}
	\write\m@ne{}
	\vbox{}
	\penalty -\@Mi
}
\makeatother

\begin{document}

\title{Multiscale modelling of nanoscaled FETs based on 2D ferroelectric materials}

\author*[1]{A. Toral-Lopez}
\email{alejandro.lopez@ing.unipi.it, damiano.marian@unipi.it}
\author[2]{M. Virgilio}
\author[1]{G. Fiori}
\author[2]{D. Marian}

\affil[1]{Dipartimento di Ingegneria dell'Informazione, Università di Pisa, Italy}
\affil[2]{Dipartimento di Fisica ``{\itshape Enrico Fermi}'', Università di Pisa, Italy}

\date{\today}

\abstract{Ferroelectric materials are highly attractive for low-power, high-speed electronics and emerging neuromorphic computing architectures. However, the severe physical scaling limits of conventional bulk (3D) ferroelectrics at the nanoscale have shifted attention towards two-dimensional (2D) ferroelectric monolayers. Accurate device-level performance predictions are essential to accelerate the experimental testing and screening of these novel materials. In this work, we present a multiscale simulation framework that bridges first-principles density functional theory with Non-Equilibrium Green's Function (NEGF) transport calculations. Using a 2D Indium Phosphide monolayer as a case study, our approach leverages a continuously interpolated, polarization-dependent Hamiltonian embedded within a self-consistent Poisson-NEGF solver. The model captures the dynamic interplay between ion movement and electronic transport, naturally reproducing macroscopic hysteresis loops and memory windows without empirical parameters. This predictive pipeline provides a computationally efficient tool to evaluate and optimize next-generation 2D ferroelectric field-effect transistors.}


\maketitle

\section{Introduction}

The demand for energy-efficient, high-speed, and high-density non-volatile memory, alongside emerging neuromorphic computing, has placed ferroelectric materials at the forefront of modern device engineering \cite{Naseer2024_InP,Naseer2025b}. These materials are defined by their spontaneous and switchable electric polarization, which can be controlled via an external electric field \cite{Naseer2025b}. While bulk (3D) ferroelectrics have long been investigated for applications such as field-effect transistors (FETs) and sensors, they face a fundamental ``thickness limit'' \cite{Junquera2003}. As these materials are scaled down, enhanced depolarization fields and surface effects often lead to a significant reduction or complete loss of ferroelectric polarization, hindering the development of ultra-scaled devices \cite{Junquera2003, Kruse2023}. Two-dimensional (2D) ferroelectric materials have emerged as a promising solution to these scaling challenges \cite{Chang2016, Fei2018, DiSante2015}. \color{black}The main feature of this technology lies in its ability to maintain robust, switchable polarization even at the monolayer limit, satisfying the ultimate vertical scaling constraints required for both ultra-dense standalone memories and synaptic elements \cite{Chang2016, Fei2018, Naseer2024_InP}. Their van der Waals nature ensures pristine, dangling-bond-free interfaces, allowing for seamless integration into flexible and low-dimensional electronics \cite{Naseer2025b}. These unique characteristics make them ideal candidates for next-generation hardware, especially for neuromorphic computing, which requires devices like memristors to emulate the synaptic plasticity of the human brain, and for highly compact memories\cite{Chua1971, Strukov2008, Ielmini2018}. \color{black}

Memristors, originally conceptualized as the missing fundamental circuit element \cite{Chua1971}, have matured into practical nanoscale devices driving neuromorphic and unconventional computing paradigms \cite{Lanza2025_Nature, Strukov2008}. Across diverse device families—including electrochemical metallization cells (EMCs), valence change mechanisms (VCM), and phase-change systems—resistive switching is inherently driven by nanoscale atomic rearrangements that reshape electronic conduction pathways \cite{Waser2009, Ielmini2018}. Recently, a major paradigm shift has moved the field from bulk oxides toward 2D layered materials (2DLMs), giving rise to the concept of ``atomristors'' \cite{Ge2018_NanoLett}. These ultra-thin devices unlock ultimate vertical scaling ($<1$~nm) and the ability to engineer defects as functional degrees of freedom \cite{Sangwan2018, Chen2020_Lanza}. However, resistive switching in these systems is governed by a complex spectrum of coexisting pathways that ranges from volatile charge trapping to structural phase transitions \cite{Zhang2019_NatMat, Farronato2022}, making them highly stochastic and strongly dependent on local atomic configurations \cite{Villena2024_Mser}. This structural sensitivity complicates empirical design rules and presents a major interpretative challenge for available theoretical methods, which often treat electron transport and ionic dynamics separately \cite{Ielmini2017, Wang2019}.

A particularly promising avenue within this landscape is the development of ultra-scaled Ferroelectric Field-Effect Transistors (FeFETs) and ferroelectric tunnel junctions (FTJs)\cite{Garcia2014}. Recent experimental results have successfully demonstrated the viability of 2D polar layers, as for example $\alpha$-In$_2$Se$_3$ or CuInP$_2$S$_6$, acting as active channels or gate elements, achieving large memory windows and high ON/OFF current ratios \cite{Si2019_FeSFET, Ghani_CIPS}. In these advanced architectures, switching the polarization state modulates the channel conductance, mitigating the severe gate leakage and severe depolarization fields that historically bottlenecked bulk-insulator FeFETs \cite{Si2019_FeSFET}. 

To accelerate the discovery and screening of these novel materials, predictive theoretical tools are indispensable \cite{Naseer2025b}. While standard \textit{ab-initio} calculations based on density functional theory (DFT) successfully predict new ferroelectric phases and static properties like spontaneous polarization ($P_s$) and energy barriers ($E_G$), bridging the gap to active, non-equilibrium device operation remains an unsolved challenge \cite{Kruse2023}. Conventional multiscale frameworks rely on extracting static DFT parameters and passing them to transport solvers, such as the Non-Equilibrium Green's Function (NEGF) formalism \cite{Datta2005,Luisier2006}. However, for authentic FeFET simulations, this static mapping is insufficient; a truly predictive framework must actively couple the ionic degrees of freedom with the electronic quantum transport under a dynamic external bias. Though recent atomistic transport models have begun addressing this, they typically treat the ferroelectric configurations as "frozen snapshots" \cite{Huang2024, Pan2023}.

In this work, we go a step forward introducing a novel multiscale approach that integrates \textit{ab-initio} DFT calculations directly into a self-consistent NEGF-Poisson scheme by explicitly coupling the structural phase transitions to the quantum transport Hamiltonian. Rather than relying on static structural approximations, our framework establishes a continuous, physical mapping from the local applied electric field to the resulting atomic displacements, macroscopic polarization, and tight-binding parameters. By fitting the extracted \textit{ab-initio} Wannier Hamiltonian parameters to analytical expressions dependent on the instantaneous polarization state, we construct an effective tight-binding model that evolves continuously. When incorporated into our \textit{in-house} quantum transport solver NanoTCAD ViDES \cite{Marian2023}, this dynamically updated Hamiltonian rigorously captures the self-consistent interplay between atomistic polarization switching and gate electrostatics on the fly.  We apply this developed methodology to III-V monolayers as a case study, with a specific emphasis on Indium Phosphide (InP) utilized in a FeFET geometry. Recent first-principles predictions indicate that InP monolayers exhibit a significantly robust out-of-plane spontaneous polarization ($P_s > 10$ pC/m) that outperforms most known 2D ferroelectrics \cite{Naseer2024_InP}. With its high predicted Curie temperature and ideal semiconducting properties, InP stands out as a promising candidate for room-temperature ferroelectric memory and neuromorphic applications \cite{Naseer2024_InP}

\section{Results and Discussion}

As mentioned in the Introduction, the core of our proposed multiscale approach lies in the direct, self-consistent coupling of the ionic degrees of freedom to electronic quantum transport. While, to the best of our knowledge, existing atomistic simulations coupled with the NEGF formalism for ferroelectric materials treat the ferroelectric lattice configurations statically \cite{Huang2024, Pan2023}, our framework establishes a continuous, physical mapping from the local applied electric field to the resulting change in atomic configuration, the macroscopic polarization, and ultimately the transport Hamiltonian itself. This dynamic pipeline allows us to evaluate device responses where the physical displacement of atoms explicitly governs electronic transport. 

The process is divided into three main parts, which are briefly summarized hereafter and detailed in the following subsections.\\ 
\textit{(i) Ab-initio Polarization Evaluation.} First, we extract the Minimum Energy Path (MEP) governing the structural ferroelectric phase transition of the material using Nudged Elastic Band (NEB) calculations \cite{Henkelman2000}. For each discrete atomic configuration along this path, the corresponding spontaneous polarization is evaluated using the Berry phase method \cite{KingSmith1993}. \\
\textit{(ii) Polarization-Dependent Wannierization.} For every configuration simulated along the MEP, we compute the electronic band structure and extract the corresponding Wannier Hamiltonian. To bridge the gap between discrete \textit{ab-initio} states and continuous device operation, we track the evolution of the onsite energies and hopping parameters along the transition path. By fitting these parameters to analytical expressions dependent on the polarization, we construct a unified tight-binding-like Hamiltonian that remains a continuous function of the polarization state.\\ 
\textit{(iii) Self-Consistent NEGF Transport.} Finally, this dynamic Hamiltonian is integrated into our in-house quantum transport simulator, NanoTCAD ViDES \cite{Marian2023}. The solver operates self-consistently, coupling the NEGF formalism with the Poisson equation. The local electric field computed across the 2D monolayer directly determines the localized atomic configuration and polarization, which in turn update the Hamiltonian for transport calculations in a closed feedback loop.

\subsection{\textit{Ab-initio} Polarization Evaluation}

We first perform first-principles DFT calculations to determine the electronic and polarization properties of 2D InP using the Quantum ESPRESSO package \cite{Giannozzi2020}; comprehensive details regarding these calculations are provided in the Methods section. In particular, the first-principles calculations include structural relaxation, self-consistent field computations, NEB evaluations, and Berry phase extractions.

\begin{figure}[th]
    \centering
        \includegraphics[width=\textwidth]{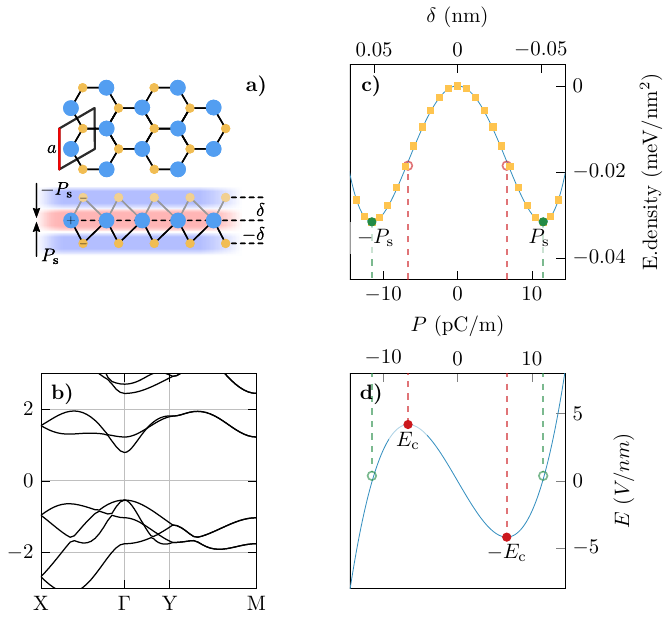}
    \caption{(a) Schematic depiction of the atomic structure and unit cell parameters of the 2D InP monolayer. (b) Electronic band structure of the minimum-energy ferroelectric configuration calculated via DFT. Although the calculations reported in this subsection were performed using the hexagonal structure, the electronic band structure reported here refers to the orthorhombic cell used for the Wannierization and the transport device calculations (see Methods for more details). (c) Energy density versus macroscopic polarization \color{black} (the units of the polarization are reported according to the Supplemental Material of Ref. \cite{DiSante2015}) \color{black}, exhibiting the characteristic double-well potential. Data points are extracted from NEB and Berry phase calculations, while the solid line represents the Landau-Ginzburg analytical fit. (d) Electric field versus polarization profile ($\mathbf{E}$-$P$ curve) derived from the derivative of the free energy. The local extrema (dots) indicate the coercive field $\mathbf{E}_{\rm c}$ required for polarization switching.}
    \label{fig:ferromatpolenergy}
\end{figure}

Figure \ref{fig:ferromatpolenergy}(a) illustrates the atomic structure of the 2D InP unit cell considered in our simulations. Our structural relaxation yields an optimized lattice constant of $a = 4.215$ \AA\ and a buckling distance between In and P atoms of $\delta = 0.512$ \AA. In its ground state, the breaking of spatial inversion symmetry gives rise to a spontaneous out-of-plane polarization of $P_s = 11.52$ pC/m, in accordance with previous results \cite{Naseer2024_InP}. The corresponding electronic band structure for this minimum-energy ferroelectric configuration is shown in Figure \ref{fig:ferromatpolenergy}(b). We find that the material is a two-dimensional semiconductor with a bandgap of $E_g = 1.34$ eV.

To characterize the polarization switching, we evaluate the MEP between the two degenerate ferroelectric ground states (upward and downward polarization) using the NEB method \cite{Henkelman2000}. In particular, the ferroelectric transition in InP is achieved through the vertical displacement of the phosphorus atom from position $-\delta$ to $+\delta$ (see Figure \ref{fig:ferromatpolenergy}(a)). The NEB calculations provide the atomic configurations at intermediate positions between the upward and downward polarization states. For each intermediate atomic configuration (snapshot) along this transition path, the macroscopic polarization $P$ is evaluated using the Berry phase approach \cite{KingSmith1993}, as indicated by the yellow square data points in Figure \ref{fig:ferromatpolenergy}(c). Mapping the system's energy density $G$ against the calculated polarization reveals the characteristic double-well energy landscape typical of ferroelectric phase transitions. To analytically capture this continuous transition for subsequent device-level simulations, the \textit{ab-initio} data are fitted to a macroscopic Landau-Ginzburg phenomenological model \cite{Landau1937}:
\begin{equation}
    G(P) = \frac{A}{2}P^{2} + \frac{B}{4}P^{4},
    \label{eq:GL}
\end{equation}
where the coefficients $A = -4.73 \cdot 10^{14}$ meV/pC$^2$ and $B = 1.79 \cdot 10^{30}$ meV$\cdot$nm$^2$/pC$^4$ are obtained by fitting Eq.~(\ref{eq:GL}) to the calculated snapshots (yellow points) using a least-squares method.

Once the analytical $G(P)$ profile is extracted, depicted by the solid blue line in Figure \ref{fig:ferromatpolenergy}(c), the relationship between the free energy and polarization allows us to derive the equation of state for the internal electric field $\mathbf{E}$ as a function of $P$ \cite{Devonshire1949}:
\begin{equation}
    \mathbf{E}(P) = \frac{\partial G}{\partial P} = A P + B P^{3}.
    \label{eq:Efield}
\end{equation}
The resulting $\mathbf{E}(P)$ profile is plotted in Figure \ref{fig:ferromatpolenergy}(d). Extracting this continuous analytical profile serves two critical purposes for our multiscale framework. First, it allows us to identify the theoretical coercive field, $\mathbf{E}_{\rm c} = 4.18$ V/nm, indicated by the red dots in Figure \ref{fig:ferromatpolenergy}(d) at the local extrema of the curve. This represents the minimum external electric field required to overcome the central energy barrier and switch the polarization state. Secondly, it establishes a direct mathematical link utilized in the self-consistent device simulations, where the polarization of the channel can be directly linked and updated in response to the local electric field applied.

\color{black} It is worth noting that while monolayer InP has been highlighted as a promising candidate due to its large spontaneous polarization, it remains an emerging material that, to the best of our knowledge, has not yet been experimentally synthesized. Our first-principles evaluation yields an intrinsic coercive field of $E_{\rm c} = 4.18$ V/nm. From a device application perspective, although this high field might not compromise the InP monolayer itself, whose intrinsic electrical breakdown limit remains unknown, it poses a technological challenge for the double-gate architecture discussed in Section \ref{Transport}. Specifically, the electrostatic potential required to flip the channel polarization induces an internal electric field across the thin insulating layers that overcomes the  dielectric breakdown threshold of conventional oxides such as SiO$_2$. Far from being a limitation of our model, this finding highlights the predictive power and utility of our self-consistent Poisson-NEGF pipeline: it successfully exposes critical bottlenecks prior to physical fabrication, proving to be a versatile CAD workflow that can be seamlessly applied to any alternative 2D ferroelectric material family. Furthermore, it must be emphasized that the coercive fields computed in this work via ideal first-principles calculations should be regarded as theoretical upper limits, representing a worst-case scenario of uniform, homogeneous switching. In realistic experimental samples, extrinsic factors such as crystal defects, edge states, or domain-wall nucleation typically lower the required switching threshold, meaning that InP could remain a viable candidate for future technology nodes \cite{Petralanda2022}.
\color{black}

\subsection{Polarization-Dependent Wannierization}

\begin{figure}[th]
    \centering
    \includegraphics[width=0.9\textwidth]{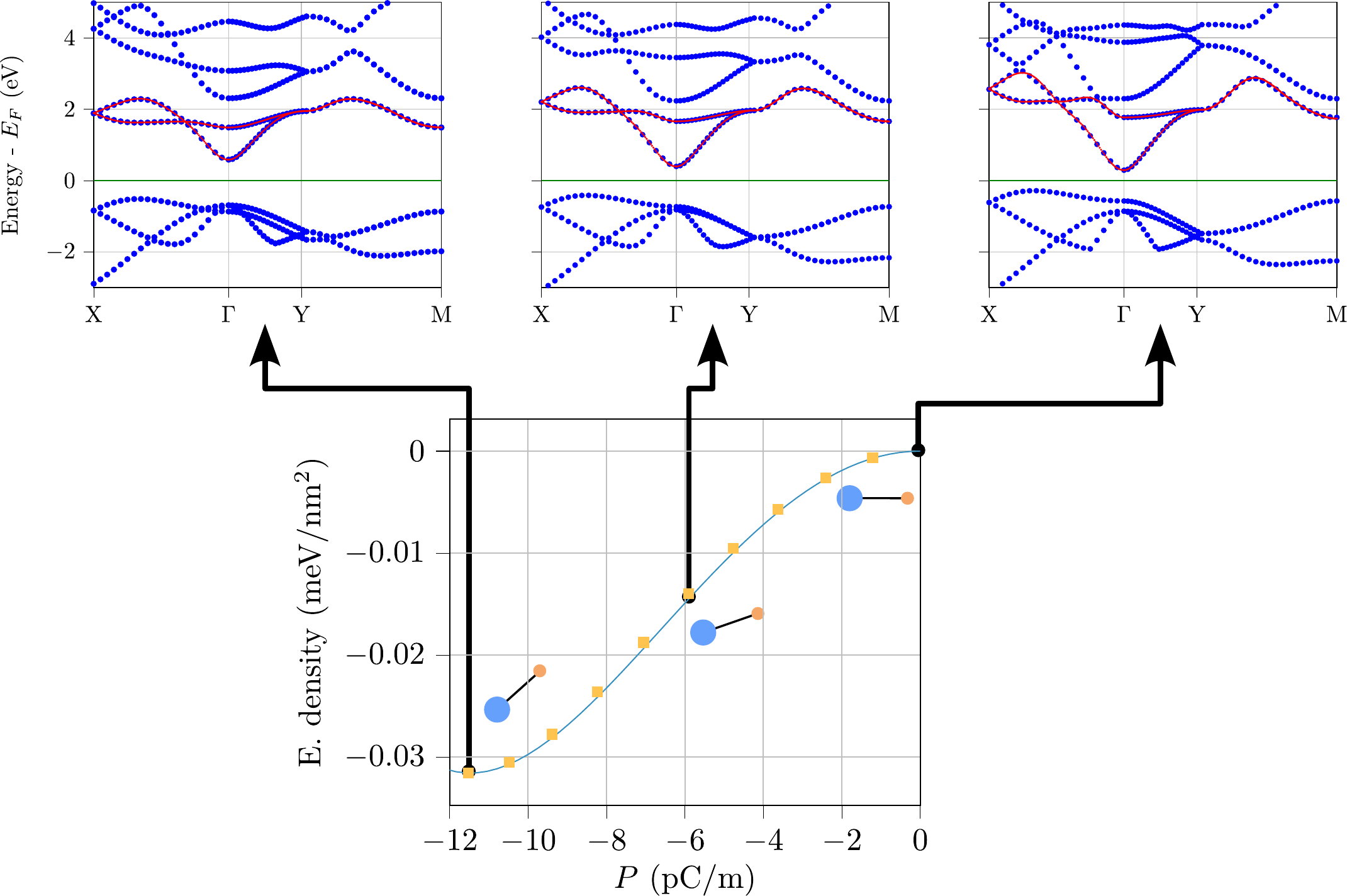}
    \caption{Evolution of the electronic band structure for three representative snapshots along the Minimum Energy Path (MEP) extracted from NEB calculations. Blue dots correspond to the first-principles bands calculated with Quantum ESPRESSO, while the red solid lines represent the bands resulting from the Wannierization process.}
    \label{fig:modellingferroelectric}
\end{figure}

To bridge the gap between discrete \textit{ab-initio} calculations and continuous device-level transport simulations, we developed a dynamic tight-binding representation of the Hamiltonian of the material. For each atomic configuration (snapshot) extracted along the MEP, yellow points in Figure \ref{fig:ferromatpolenergy}(b), the corresponding electronic Hamiltonian, which Quantum ESPRESSO provides in a plane-wave basis, was processed using the Wannier90 code \cite{Mostofi2014} to construct a Hamiltonian based on Maximally Localized Wannier Functions (MLWFs). Comprehensive details of the Wannier calculations are reported in the Methods section. Figure \ref{fig:modellingferroelectric} presents the resulting band structures for three distinct structural configurations spanning the first half of the polarization path (due to symmetry the other half is equivalent). The excellent agreement between the \textit{ab-initio} DFT bands (blue dots) and the Wannier-interpolated bands (red lines) confirms that the extracted discrete Hamiltonians accurately capture the evolution of the electronic characteristics as the lattice undergoes its ferroelectric transition.

While these discrete Hamiltonians perfectly describe the electronic states at specific points along the MEP, realistic device simulations require a continuous Hamiltonian that responds to an arbitrary local electric field. To achieve this, we utilized the extracted dataset to construct a polarization-dependent tight-binding model. First, we leveraged the crystal symmetries to group equivalent hopping parameters, thus narrowing our focus solely to the inequivalent tight-binding parameters. We then systematically tracked the evolution of these reduced hopping parameters ($t_{ij}$) and onsite energies ($\varepsilon_i$) across all simulated configurations along the MEP.

By mapping these extracted matrix elements against the macroscopic polarization $P$ calculated in the previous step, we interpolated their discrete values using cubic splines to construct continuous analytical functions for the onsite energies $\varepsilon_i(P)$ and hopping parameters $t_{ij}(P)$. Consequently, the discrete tight-binding models are unified into a single, comprehensive continuous Hamiltonian operator $\hat{H}(P)$:
\begin{equation}
    \hat{H}(P) = \sum_{i} \varepsilon_i(P) \hat{c}^\dagger_i \hat{c}_i + \sum_{\langle i,j \rangle} t_{ij}(P) \hat{c}^\dagger_i \hat{c}_j + \sum_{\langle\langle i,j \rangle\rangle} t'_{ij}(P) \hat{c}^\dagger_i \hat{c}_j + \dots,
    \label{eq:HP}
\end{equation}
where $\hat{c}^\dagger_i$ and $\hat{c}_i$ are the creation and annihilation operators for an electron in the maximally localized Wannier state at site $i$, and the summations explicitly run over nearest neighbors ($\langle i,j \rangle$), next-nearest neighbors ($\langle\langle i,j \rangle\rangle$), and all subsequent long-range interactions captured by the Wannierization. By utilizing Eq. (\ref{eq:Efield}) the polarization can be evaluated as a function of the local electric field, i.e., $P = P(\mathbf{E})$. This allows the formulation in Eq. (\ref{eq:HP}) to naturally translate into a dynamic, electric field-dependent Hamiltonian:
\begin{equation}
    \hat{H}(\mathbf{E}) = \hat{H}\Big(P(\mathbf{E})\Big).
    \label{eq:HE}
\end{equation}
This continuous analytical formulation, which preserves the full complexity of the \textit{ab-initio} electronic structure, is the cornerstone of our multiscale methodology.

\begin{figure}[th]
    \centering
    \if pdfFigures
        \input{Figures/InterpBands.tex}
    \else
        \includegraphics[width=\textwidth]{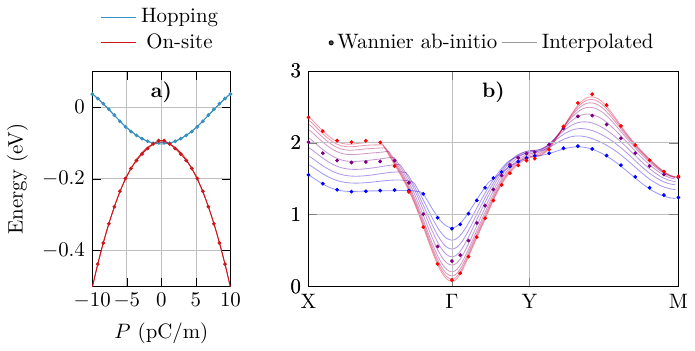}
    \fi
    \caption{a) Representative Wannier on-site energy ($\varepsilon$) and first neighbor hopping parameter ($t_{ij}$) as a function of the macroscopic polarization, comparing discrete \textit{ab-initio} data (symbols) against the analytical fitting (lines). b) Evolution of the lowest conduction bands as a function of the macroscopic polarization. The marks denote the eigenvalues extracted directly from the discrete Wannier Hamiltonians for specific \textit{ab-initio} structural snapshots. The solid lines represent the band structures generated by the continuous, interpolated Hamiltonian $H(P)$. The model perfectly captures the band topology at the explicitly sampled configurations while also predicting the smooth electronic evolution for intermediate, unsampled polarization states.}
    \label{fig:interpbands}
\end{figure}

Figure \ref{fig:interpbands} illustrates the robustness and predictive power of our continuous, interpolated $H(P)$ model. First, Figure \ref{fig:interpbands}(a) reports a representative on-site energy ($\varepsilon$) and a hopping parameter ($t_{ij}$) as a function of the polarization, comparing the values obtained from discrete snapshots (symbols) against the analytical fitting. Second, Figure \ref{fig:interpbands}(b) compares the lowest eigenvalues of the conduction band obtained directly from the discrete Wannier Hamiltonians (markers) against those generated by our analytical $H(P)$ model (solid lines) for the explicitly sampled \textit{ab-initio} configurations. The excellent agreement in these discrete points validates our fitting procedure, demonstrating that the interpolated Hamiltonian preserves the precise electronic band. Crucially, the figure also displays the band structures generated by $H(P)$ for several intermediate polarization states where no \textit{ab-initio} data were explicitly computed. The smooth, physical transition of the energy bands across these intermediate states highlights the model's ability to continuously describe the material's electronic response to structural deformations. This capability is fundamental for self-consistent device simulations: it provides an uninterrupted evolution of the electronic structure under a continuously varying electric field, drastically reducing the computational overhead by eliminating the need for dense DFT sampling along the structural transition path.

\subsection{Self-Consistent NEGF Transport}
\label{Transport}

\begin{figure}[th!!!]
    \centering
        \includegraphics[width=\textwidth]{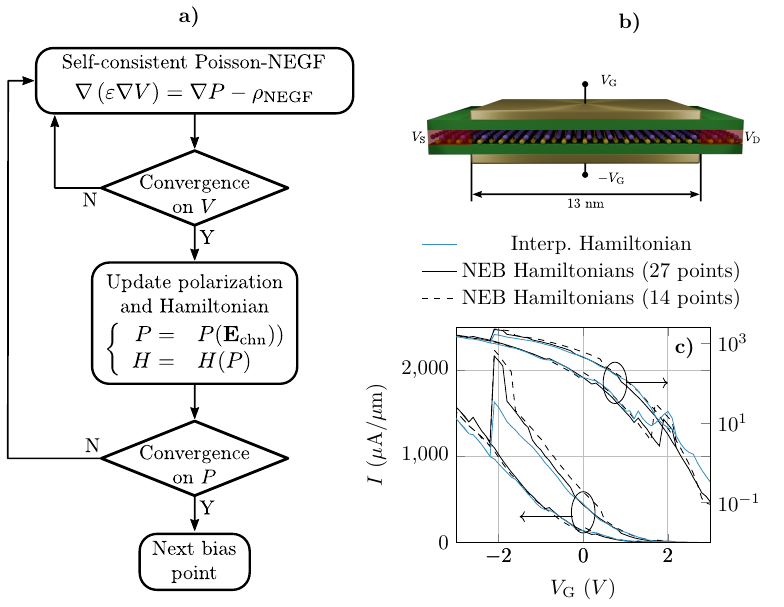}
    \caption{Overview of the multiscale device simulation framework. a) Flowchart of the modified self-consistent Poisson-NEGF transport solver, highlighting the inner convergence loop on the electrostatic potential $V$ and the outer feedback loop on the ferroelectric polarization $P$ and polarization-dependent Hamiltonian $H(P)$. b) 3D schematic representation of the simulated double-gated InP FeFET with a 13~nm channel length under asymmetric gate biasing ($\pm V_{\text{G}}/2$). c) Simulated transfer characteristics ($I_{\text{DS}} - V_{\text{G}}$) comparing, interpolated Hamiltonian model (blue solid line) against the discrete NEB snapshot Hamiltonians sampled at 27 points (black solid line) and 14 points (dashed line).}
    \label{fig:curvesidsvgs1}
\end{figure}

To bridge the gap between microscopic material properties and macroscopic device performance, the continuous, polarization-dependent Hamiltonian model is embedded into a quantum transport framework driven by the open-source simulator NanoTCAD ViDES \cite{Marian2023}. To capture the dynamic, self-consistent coupling between the external electrostatic bias and the internal ferroelectric state of the monolayer, the conventional Poisson-NEGF (Non-Equilibrium Green's Function) scheme has been augmented. Specifically, the electrostatic solver is reformulated to explicitly incorporate the bound charge density generated by the channel's spontaneous polarization. Given that the 2D ferroelectric monolayer features an out-of-plane polarization vector oriented strictly along the transport confinement axis ($y$-axis, as illustrated in the device schematic in Figure \ref{fig:curvesidsvgs1}(b)), the modified Poisson equation is expressed as: 
\begin{equation}
    \nabla \cdot \left( \varepsilon \nabla V \right) = -\rho - \frac{\partial P_{\rm y}}{\partial y},
\end{equation}
where $V$ is the electrostatic potential, $\varepsilon$ is the local dielectric permittivity, $\rho$ is the free mobile charge density computed via NEGF, and $P_{\rm y}$ is the out-of-plane macroscopic polarization. 

The integration of this modified Poisson equation allows the electrostatic potential to drive the self-consistency in two simultaneous ways: (i) updating the local polarization state of the material, and (ii) dynamically updating the Hamiltonian utilized for the NEGF transport calculations. At each iterative step, the local transverse electric field $\mathbf{E}$ is evaluated from the potential profile. Using the relationship $\mathbf{E}(P)$ in Eq. (\ref{eq:Efield}), the simulator determines the updated localized polarization $P_{\rm y}$. This value is then fed directly into our formulation $\hat{H}(P)$ to construct a newly updated tight-binding Hamiltonian. The NEGF solver utilizes this configuration-accurate Hamiltonian to compute the quantum carrier transport, generating a new charge density $\rho$ that is passed back to the Poisson solver. The flowchart of the simulator is reported in Figure \ref{fig:curvesidsvgs1}(a). 

To validate these model enhancements, we simulated an ultra-scaled InP-based Ferroelectric Field-Effect Transistor (FeFET). The simulated device features a 13 nm long active channel with 35 nm long highly doped source and drain regions. The device is modulated by a double-gate architecture (top and bottom), with each gate separated from the channel by a 0.5 nm thick oxide layer (see Figure~\ref{fig:curvesidsvgs1}(b)). This dual-gate structure, combined with asymmetric biasing, provides the robust control over the internal transverse electric field required to switch the polarization state of the InP monolayer.

\begin{figure}[th]
    \centering

        \includegraphics[width=\textwidth]{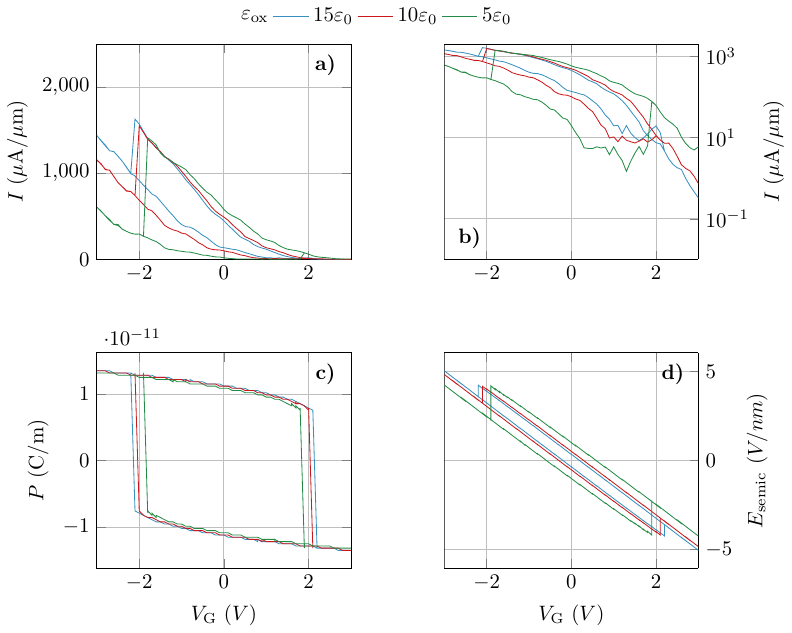}
    \caption{Impact of gate oxide permittivity ($\varepsilon_{\text{ox}} \in \{15\varepsilon_0, 10\varepsilon_0, 5\varepsilon_0\}$) on the simulated InP FeFET performance. The top panels display the device transfer characteristics in linear (a) and logarithmic (b) scales for dielectric constants of $15\varepsilon_0$ (blue), $10\varepsilon_0$ (red), and $5\varepsilon_0$ (green). The bottom panels illustrate the corresponding hysteresis loops of the macroscopic polarization $P$ (c) and the electric field $E_{{\rm semic}}$ (d) across the 2D semiconductor channel layer as a function of the applied gate voltage ($V_{\text{G}}$).} 
    
\label{fig:curvesidsvgs}
\end{figure}

The transfer curves of the drain-source current ($I_{\rm DS}$) as a function of the gate voltage ($V_{\rm G}$) depicted in Figure \ref{fig:curvesidsvgs1}(c) highlight the capabilities of our dynamic Hamiltonian approach. \color{black} In particular, the memory window and the corresponding hysteresis loop emerge completely naturally, exhibiting a sharp transition at $V_{\rm G} \approx \pm 2$~V. This total voltage window of $\Delta V \approx 4$~V, applied across a thickness of approximately $1.2$~nm, generates an internal electric field, $E_{{\rm semic}}$ across the 2D InP semiconductor (see also Figure~\ref{fig:curvesidsvgs}d), that directly correlates with the coercive field of $4.18$~V/nm previously derived from our \textit{ab-initio} calculations. \color{black} It is important to emphasize that dynamically updating the Hamiltonian based on the local electrostatic environment, even when utilizing discrete \textit{ab-initio} snapshots (black line), represents a substantial advancement in the device-level simulation of ferroelectrics. However, utilizing the continuous, interpolated $H(P)$ approach (blue line) provides a remarkably smoother and more physically representative transition profile during polarization reversal, as it avoids the abrupt discontinuities inherent to discrete sampling. Furthermore, as observed at the polarization extremes, the asymptotic current values differ between the two methods; the interpolated Hamiltonian captures the continuous evolution of the electronic band structure more comprehensively, leading to distinct current levels at both high and low gate biases compared to its discretely sampled counterpart.

Leveraging this dynamical framework, we investigated the influence of gate electrostatics on the polarization switching by varying the permittivity of the insulating oxide layers ($\varepsilon_{\text{ox}} \in \{15\varepsilon_0, 10\varepsilon_0, 5\varepsilon_0\}$), with $\varepsilon_0$ the vacuum permittivity, as depicted in Figure~\ref{fig:curvesidsvgs}. The simulations reveal a distinct physical trend: reduced dielectric constants weaken the electrostatic coupling between the gate and the channel. Consequently, the bound charges associated with the ferroelectric polarization dominate the local internal electric field, significantly reshaping the channel's electrostatic profile. This degraded gate coupling dictates that a larger external voltage is required to overcome the material's coercive field, thereby expanding the width of the hysteresis loop. Crucially, the memory windows observed in these transfer characteristics are not the product of empirical fitting parameters. Instead, they emerge intrinsically from our multiscale methodology. The rigorous self-consistent loop, i.e. coupling the external bias, the uncompensated bound charges, and the dynamically updated quantum Hamiltonian, naturally yields macroscopic memory effects directly from first principles.

\section{Conclusions}
In this work, we have presented a rigorous multiscale simulation framework for two-dimensional ferroelectric field-effect transistors that treats structural ionic reconfigurations and quantum electronic transport on an equal footing. By bridging first-principles minimum energy paths with a self-consistent Poisson-NEGF transport solver, our approach successfully captures the dynamic evolution of a FeFET based on 2D ferroelectric material (InP) under external bias. 

The core methodological advancement lies in the development of a compressed, polarization-dependent Hamiltonian. Through an efficient symmetry-based interpolation scheme, the framework bypasses the need for computationally prohibitive, discrete ab-initio snapshots, ensuring a physically smooth and continuous description of the switching dynamics. Crucially, the macroscopic memory windows and hysteresis loops emerge natively from the coupled electrostatic-structural feedback loop rather than from empirical phenomenological parameters. This purely predictive pipeline offers a robust and computationally viable tool for the direct screening and optimization of low-dimensional non-volatile memory architectures and neuromorphic hardware.

\section*{Methods}
\label{sec:methods}

\subsection*{\textit{Ab-initio} Density Functional Theory Calculations}

The first-principles calculations were performed within the framework of Density Functional Theory (DFT) as implemented in the open-source \textsc{Quantum ESPRESSO} distribution \cite{Giannozzi2020}. The exchange-correlation interaction was described using the Generalized Gradient Approximation (GGA) parametrized by the Perdew-Burke-Ernzerhof (PBE) functional. The electron-ion interactions were modeled using Projector Augmented-Wave (PAW) pseudopotentials from the \textit{pslibrary}, explicitly treating the semi-core $d$-electrons of Indium.

The electronic wavefunctions and the charge density were expanded using plane-wave basis sets with kinetic energy cutoffs of 60~Ry and 480~Ry, respectively. To simulate the 2D InP monolayer and prevent spurious interactions between periodic periodic images along the out-of-plane direction, a large unit cell parameter of $c = 15.495$~\AA\ was utilized. For the integration over the Brillouin zone, a $\Gamma$-centered $21 \times 21 \times 1$ Monkhorst-Pack $k$-point mesh was employed. 
The structural optimization was carried out until the total energy variations dropped below $2 \times 10^{-5}$~Ry and the Hellmann-Feynman forces on all atoms were less than $1 \times 10^{-4}$~Ry/bohr. The self-consistent field (SCF) calculations were iterated until an energy convergence threshold of $4 \times 10^{-10}$~Ry was achieved.

To evaluate the macroscopic spontaneous polarization of the material, we employed the modern theory of polarization using the Berry phase approach. 
Finally, to investigate the transition mechanism and the kinetic energy barrier associated with the polarization reversal from the $+P_{\rm z}$ to the $-P_{\rm z}$ state, we performed Climbing Image Nudged Elastic Band (CI-NEB) calculations. The minimum energy path (MEP) was discretized using 21 or 14 intermediate images interpolated between the two fully relaxed polarization ground states. The atomic positions along the transition path were evaluated with a $15 \times 15 \times 1$ $k$-point grid and without enforcing symmetry constraints.

\subsection*{Wannier Tight-Binding Hamiltonian Extraction}

To obtain Hamiltonians based Maximally Localized Wannier Functions (MLWFs)  we have used \textsc{Wannier90} code interfaced with \textsc{Quantum ESPRESSO}. Because device-level Non-Equilibrium Green's Function (NEGF) transport requires an orthogonal grid aligned with the transport and transverse directions, the relaxed primitive hexagonal unit cell was mapped onto an equivalent rectangular supercell. This rectangular cell contains 4 atoms (two Indium and two Phosphorous) with lattice parameters of $a = 4.215$~\AA\ and $b = 7.300$~\AA.

For each structural image along the polarization switching path (the Minimum Energy Path), a fully self-consistent field (SCF) calculation was performed on the rectangular cell using an $11 \times 11 \times 1$ Monkhorst-Pack $k$-point mesh. 

\section*{Declarations}

\subsection*{Availability of data and material}
The data used and analysed during the current study are available from the corresponding author on reasonable request.

\subsection*{Competing interests}
The authors declare no competing interests

\subsection*{Funding}
European Union’s Horizon Europe: CHIPS-JU, contract No. 101194458 (ENERGIZE)

\subsection*{Authors' contribution}
All of the authors contributed equally to the conceptualization, methodology and writing of the study. A. Toral-Lopez and D. Marian contributed to the data curation, formal analysis and software development. G. Fiori, D. Marian and M. Virgilio contributed to the supervision of the work. G. Fiori and D. Marian contributed to the founding and project administration.

\subsection*{Acknowledgements}
We acknowledge funding from the European Union’s Horizon Europe research and innovation program: via CHIPS-JU, under the project ENERGIZE (101194458)

\phantomsection 
\label{Bibliogra}
\bibliography{mrt}

\printglossary[type=\acronymtype]

\end{document}